# Advanced DAG-Based Ranking (ADR) Protocol for Blockchain Scalability

**Tayyaba Noreen[1,*], Qiufen Xia[1] and Muhammad Zeeshan Haider[2]**

[1]International School of Information Science and Engineering, Dalian University of Technology, Dalian, 116620, China
[2]School of Software, Dalian University of Technology, Dalian, 116620, China
*Corresponding Author: Tayyaba Noreen. Email: tayyabanoor849@gmail.com

**Abstract:** In the past decade, blockchain has evolved as a promising solution to develop secure distributed ledgers and has gained massive attention. However, current blockchain systems face the problems of limited throughput, poor scalability, and high latency. Due to the failure of consensus algorithms in managing nodes' identities, blockchain technology is considered inappropriate for many applications, e.g., in IoT environments, because of poor scalability. This paper proposes a blockchain consensus mechanism called the Advanced DAG-based Ranking (ADR) protocol to improve blockchain scalability and throughput. The ADR protocol uses the directed acyclic graph ledger, where nodes are placed according to their ranking positions in the graph. It allows honest nodes to use the Direct Acyclic Graph (DAG) topology to write blocks and verify transactions instead of a chain of blocks. By using a three-step strategy, this protocol ensures that the system is secured against double-spending attacks and allows for higher throughput and scalability. The first step involves the safe entry of nodes into the system by verifying their private and public keys. The next step involves developing an advanced DAG ledger so nodes can start block production and verify transactions. In the third step, a ranking algorithm is developed to separate the nodes created by attackers. After eliminating attacker nodes, the nodes are ranked according to their performance in the system, and true nodes are arranged in blocks in topological order. As a result, the ADR protocol is suitable for applications in the Internet of Things (IoT). We evaluated ADR on EC2 clusters with more than 100 nodes and achieved better transaction throughput and liveness of the network while adding malicious nodes. Based on the simulation results, this research determined that the transaction's performance was significantly improved over blockchains like Internet of Things Applications (IOTA) and ByteBall.

**Keywords:** Blockchain; scalability; directed acyclic graph; advanced DAG-based ranking protocol

## 1 Introduction

As a distributed system with no centralized authority, blockchain has recently gained interest from both industry and academia. With its public nature, decentralization, distributed structure, and unique characteristics, blockchain technology is bringing innovative changes to every aspect of life [1–3]. Network participants share the same ledger, which makes it a reliable and secure system. Satoshi Nakamoto initially proposed the idea of a distributed ledger [4] in his paper, which was then developed into Bitcoin. Bitcoin was introduced as the first blockchain technology in which blocks are arranged in a chain. To maintain the chain, Bitcoin uses the Proof of Work (PoW) consensus algorithm, in which miners try to solve cryptographic puzzles and get coins as a reward. Bitcoin and Namecoin are among the blockchain applications that depend on coins to motivate miners to create blocks [5]. Using these applications, non-payment of miners would compromise blockchain security [6]. Moreover, scalability is one of the most challenging aspects of making blockchain technology suitable for mainstream usage. For example, all transactions must be managed by each node on the blockchain network. All the data of the entire network and the processing capacity of the whole blockchain system are limited by the processing capacity of a single computing node. Furthermore, the growth of nodes affects the consensus algorithm, which results in an overall decrease in the system's processing capacity [7]. In existing blockchain systems, Bitcoin's consensus algorithm (PoW) processes 7 transactions per second (TPS). In contrast, Ethereum uses the Proof of Stake (PoS) consensus algorithm [8], which processes 20 transactions per second. The Hyperledger Blockchain uses Practical Byzantine Fault Tolerance (PBFT) [9], which handles 3,500 transactions per second, while Visa supports over 4500 transactions per second. The low throughput of traditional blockchains is incapable of meeting the demands of large-scale trading scenarios [10]. Researchers have proposed various solutions to address this issue; among these solutions, DAG-based IOTA [11] allows multiple blocks to add tails simultaneously, striking a balance between speed and asynchronization. As more blocks are added to the network, the security of IOTA decreases, and the confirmation time of transactions becomes unpredictable. Even though several blockchain solutions have been proposed, such as blockDAG [12], Phantom [13], and Spectre [14], that tried to improve security and performance, but their data structures were complicated, and blockchain users prefer simple and deterministic data structures.

In response to the above discussed problems, we propose an Advanced DAG-based Ranking (ADR) protocol. It consists of a directed acyclic graph block structure. Here, a block in DAG is the same as a block in a blockchain. It also comprises the block header and transactions. In contrast, blockchain blocks are always referenced by the previous block, while multiple preceding blocks can reference DAG blocks. Ranking can be defined as the rating of a member's trustworthiness by others. To create a positive interaction, ranking systems are applied so participants can trust one another [15].

Ranking and DAG make several contributions: ranking serves as an incentive, and DAG ensures the security of records and increases the scalability of the blockchain system. In ADR, a decentralized protocol is proposed that is capable of recording transactions without the involvement of any central third parties. The protocol is used in the private blockchain, where an access control layer is built into the blockchain nodes. Participants use asymmetric cryptography to authenticate each other [16]. The reason behind selecting a private blockchain is that ranking depends on identity and the time taken to accumulate it. Every participant in the ADR protocol maintains a distributed ledger containing ranking evidence, which achieves consistency with the consensus algorithm. In comparison to PoW, since no miners or hashing power is utilized during block competitions, the ADR protocol uses ranking as an incentive, which is more cost-efficient. Due to blockchain's cryptographic properties, ranking evidence can be more secure and reliable. The following is a general summary of this research.

- This research proposes an improved DAG-based algorithm for IoT blockchain, termed ADR. Through a deposit penalty mechanism, the protocol secures node reliability and increases blockchain security by using timestamps and reputation data from IoT devices. The behavior of the node and its reputation value are closely related. It can be utilized as a reward system to motivate IoT devices to act as consensus nodes and exhibit good behavior.

- The security layers were stated in the directed acyclic graphs (DAG), and the ranking of nodes was defined in combination with the DAG, making the system more scalable and stable. The ADR's random election mechanism ensures impartiality and randomness when choosing nodes for the IoT blockchain system.

- The Session Control Protocol (SCP), a dynamic join and departure verification mechanism lacking the liveness of a network in traditional blockchains, is also included in ADR. By using this approach, blockchain systems may quickly agree and act decisively in reaction to changing network conditions. This approach enables private key identity verification as well as a complete entry and exit consensus mechanism with exit and entry tickets.

- In this paper, a blockchain prototype using the ADR was developed, and extensive testing was conducted with approximately 30–40 nodes, some of which served as simulated server workstations for IoT devices. In terms of throughput, consensus effectiveness, and network communication overhead, the proposed ADR protocol shows comprehensive improvement compared to methods like IOTA and ByteBall.

The remainder of this paper is organized in the following sequence: Section 2 discusses the previous work. The ADR protocol design and implementation are presented in Section 3, and the experiment results are discussed in Section 4. Finally, the research work is concluded in Section 5.

## 2 Related Work

The dominance of Bitcoin in cryptocurrency has exposed the scalability issues of blockchain [17]. Several studies have been conducted to address the issue of blockchain scalability. Scalability key metrics have been defined, which include throughput, latency, cost per confirmed transaction (CPCT), and bootstrapping time [18]. Throughput and latency are critical performance metrics to provide a quality user experience. However, transaction throughput is more important than other key metrics. Bitcoin has a maximum throughput of 7 TPS, while Visa can support over 4000 TPS. Bitcoin's low throughput is incapable of meeting the demands of large-scale trading scenarios. Blockchain technology was developed and is mainly used for cryptocurrency-related applications (e.g., Bitcoin, Ethereum, etc.). Blockchain technology has been incorporated into business networks through Hyperledger. Scalability problems with Fabric v0.6 for public use have been somewhat resolved by Fabric v1.0 [19]. Due to peers' ability to process transactions concurrently, the transaction pace can exceed 104 TPS. This increase can be sufficient in some application settings. However, the sheer volume of blockchain applications necessitates a considerably greater transaction rate. Furthermore, there is still unresolved research on the ledger's storage needs. Therefore, suitable schemes must be carefully designed to maximize throughput to maintain the large volume of real-world transactions. The current strategies for improving blockchain scalability performance include; increasing block capacity and generation speed, using a chain-down channel, improving the consensus algorithm, and optimizing the data structure.

The protocols designed to improve the blockchain's throughput can be categorized as on-chain and off-chain solutions. On-chain solutions focus on blockchain design, including block structures, consensus algorithms, and the main chain structure. Off-chain solutions aim to alleviate the burden of the main chain. On-chain solutions such as Bitcoin-Cash [20], compact block rely [21], sharding techniques [22–25], and various improved consensus algorithms [26–28], increase transaction throughput and scalability while decreasing latency. Off-chain solutions like Lightning Networks [29], FireLedger [30], and Sidechains [31] are still under development. The lightning network enhances Bitcoin's performance by creating micro-payment channels between nodes. Pegged side chains [31] can interact with the main chain via a two-way peg, allowing maximum flexibility.

A revised blockchain structure, called DAG, is proposed to improve blockchain throughput performance and scalability [32]. DAG allows simultaneous block generation, which means it can connect multiple blocks to the previous blocks and add more transactions to the system. Lewenberg et al. [33] used the idea of a directed acyclic graph of blocks (blockDAG) in their protocol. In this protocol, a new block references multiple previous blocks. An inclusive rule is proposed to select the main chain of the newly formed DAG. Moreover, off-chain blocks are included in the ledger if their contents do not conflict with those of previous blocks. With this proposed protocol, the system can achieve higher throughput scalability.

Among the DAG-based solutions, IOTA's DAG data structure allows the network to scale easily. Everyone participates in reaching a consensus, so the more users there are, the faster the network grows. Meanwhile, IOTA has no miners [34], and users can make transactions for free. However, as more blocks are added to the network, IOTA's security and confirmation times are not guaranteed. Besides, IOTA has a central point of failure. Some improved forking solutions have been introduced to solve the issues in IOTA. Byteball maintains the chain, which is operated by 12 witnesses verifying each transaction [35]. These nodes are trusted by the developers but compromise decentralization. Spectre is a consensus protocol that provides high throughput and faster confirmation times [14]. It adopts a voting mechanism that aims to establish a system that allows concurrent block creation. However, it cannot be extended to support the complete linear order of overall transactions. Phantom solves the issue of Spectre by providing a linear order over the blocks of DAG [13]. Phantom's linear ordering costs a lot, and confirmation times are slower than Spectre. Conflux is a more scalable and fast DAG-based blockchain system [36]. It introduced two types of edges between blocks: parent and reference edges. It maintains the pivoting chain formed by the parent edge to keep the entire order of the blockchain, but its confirmation time is still nondeterministic.

## 3 ADR Protocol

This section describes the essential components of the ADR protocol in detail.

### 3.1 Protocol Overview

The overall process of the ADR protocol, which applies to the private blockchain system, is given in Fig. 1. It helps private business systems overcome storage and scalability issues. The ADR consensus protocol comprises a DAG with blocks and nodes ranked according to their behavior to participate in the consensus mechanism, improving the blockchain system's performance. It comprises node evaluation and verification, wallet creation, the DAG ordering protocol, the node rank protocol, ranking growth evaluation, and the Session Control Protocol (SCP).

- The ADR protocol structures the block in a better way to boost scalability using the improved DAG structure; with a high degree of likelihood, blocks that were mined honestly belong to this structure.

- This research applies the natural partial ranking of the consensus to a topological ordering by designing a ranking protocol that rewards blocks inside the selected domain and penalizes blocks outside or inside the selected domain using SCP semantics.
- Nodes are encouraged to perform better and build their reputation by ranking higher while mining blocks and verifying transactions. Transactions within a block are arranged by the order in which they appear; order over blocks results in order over transactions.
- With a defined set of rules, the SCP allows nodes to enter or exit the system safely. Block mining and transaction verification tasks can be carried out by nodes with valid public and private keys, participating in the consensus reputation system, and being trustworthy nodes.

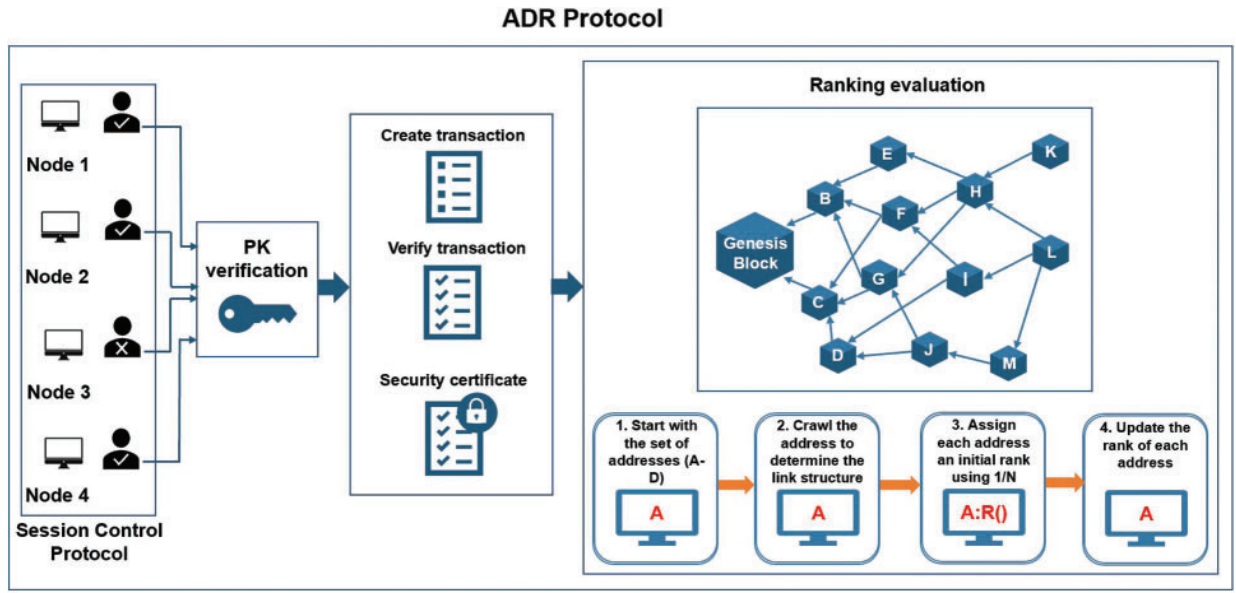

**Figure 1:** Overall architecture of ADR protocol

This research work introduces a blockchain mechanism that allows blocks to relate to each other by implementing a DAG network structure. The proposed consensus protocol is shown in Fig. 1. Node 1, Node 2, and Node 4 are the true nodes with stable ranks, while Node 3 is a faulty node with an unstable or negative rank. After the evaluation, they can participate in the consensus mechanism by mining blocks and verifying transactions.

In the ADR consensus mechanism, as given in Eq. (1), G denotes a point set V and an edge set E. Each element in the point set corresponds to a block unit that can be instantiated as a transaction Tx ∈ T, a block B ∈ B, or a layer L∈L in the protocol, where T, B, L, and R represent the sets of elements. The entry in the edge set that indicates the partial order relation between the two points a and b is a tuple (a, b). In most cases, the relationship shows that one of the block units directly or indirectly refers to another block unit. As an illustration, the notation is:

a←b means b confirms a, where {a, b} ∈ V.

$$G = (E,V)/(ul/sl) \tag{1}$$

Where,

V = {a |a ∈ R ∪ B ∪ L ∪ T},

E = {(a, b) | a←b ∧ {a, b} ∈ V}

To ensure the proper placement and collection of high-quality consensus nodes, SCP evaluates the node's entry into the system by verifying their private and public keys and creating a wallet account for true nodes. The default ranking value is given to the node, so it starts producing blocks and verifying transactions. The ADR protocol maintains the blocks in an acyclic graph structure according to the block hash algorithm, followed by the nodeRank ranking algorithm to maintain and update the ranking table of the nodes. ADR's consensus mechanism removes the faulty nodes in the system and prioritizes the true node for joining consensus according to the rank and performance of the nodes. The faulty nodes with negative ranks will be discarded and given an exit ticket by the SCP to avoid any attack on the system by faulty nodes. It makes the system secure and enhances the dynamics of the blockchain system. Blockchain networks are generally more scalable with the ADR protocol because it increases node reliability and security.

### 3.2 Node Evaluation and Verification

The ADR consensus protocol is composed of an n-number of nodes. Each node has an updated ranking value and a ranking growth rate, which determines whether the node can become a consensus node given certain conditions. If the node shows unstable behavior or is assigned a negative rank, then the node is not eligible to participate in the consensus process. The Session Control Protocol (SCP) will be activated, and a faulty node will be sent out of the system by assigning an exit ticket to the node. There is a process of node evaluation and verification that includes the following points:

- **Node wallet account:** Consensus verifies the node wallet account by verifying public and private keys, and a default rank is given to the node.
- **Node health:** Nodes create different transactions and produce blocks that ultimately update their rank in the ranking table, which is directly proportional to the node's health.
- **Ranking growth:** For maintaining the ranking of the node, the nodeRank algorithm provides a fair way to keep the ranking score of nodes based upon their fair mining of blocks and transactions.
- **Waiting nodes:** These are a group of nodes that are rapidly replaced when consensus nodes fail or exit.
- **Ranking record:** A local metadata pool is created by each node to store all nodes' behavior in a table. This table updates after each transaction is broadcast to all other nodes.

### 3.3 ADR Block Structure

The ADR consensus mechanism proposes an advanced DAG architecture to overcome the limitations of the traditional consensus mechanism. The ADR block structure in Fig. 2 describes the block event with parameters. In a DAG consensus mechanism, nodes can add their blocks to the blockchain at any time as long as they have processed the previous transactions. This way, many branches are created simultaneously, a process called forking. Forking is usually considered an issue in many traditional consensus mechanisms since it would cause the problem of "double spending". The ADR develops novel protocols and algorithms to overcome the double spending issue. It also prevents new arrival transactions from accessing the blockchain network in a forking topology. As a result, the TPS and confirmation rate will no longer be constrained. Furthermore, while massive forking blocks secure the data held in the DAG, a single user only requires a small number of resources to generate a new block. As a result, the professional miner disappears, and transaction fees might be low or zero, which is crucial for the blockchain ecosystem.

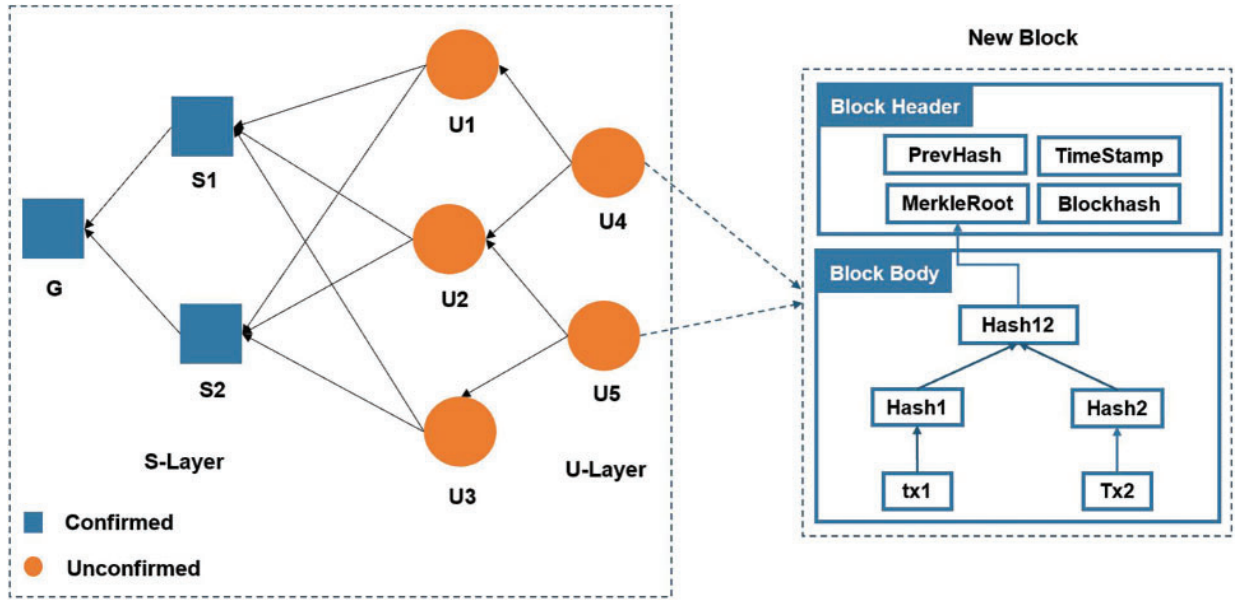

**Figure 2:** Enhanced block structure

Like traditional blockchains, the ADR blockchain allows the new transaction to reach consensus without the assistance of a miner. It aims to address the vast number of micro-transactions in IoT systems and requires that the new transaction approve the previous transactions as soon as possible. Thus, any vertex in an acyclic graph can be viewed as a block that only records one transaction. A block in the DAG also has a block header that comprises the Prevhash, timestamp, Merkleroot, and block hash and contains several transactions, just as a block in the traditional blockchain does. In block mining, the miner uses an initial random number to try to solve a difficult computation task. The sequence of transactions is determined by which the miner completes the task first and gets to produce the subsequent block in the chain. This mechanism of the network reaching consensus on the order of transactions is implemented by most traditional blockchains. Each block in a blockchain always refers to one previous block. However, blocks inside an ADR can refer to several previous blocks. The ADR protocol handles more transactions in each period than the other connected blockchains. There is more storage capacity in the ADR compared to the IOTA and ByteBall blockchains. The ADR protocol offers a promising solution for scalability in decentralized networks by adding two-dimensionality to the one-dimensional data structure of blockchains.

The transactions made during a given period are recorded into blocks, which are the basis of the blockchain network. A block stores information; each consists of several pieces of information but does not take up a lot of storage space. A block in ADR consists of the following parameters:

- **BlockHeader:** A block header identifies a specific block on a blockchain and is hashed repeatedly by nodes to gain mining ranking. A blockchain consists of various blocks that store information regarding transactions on a blockchain network. Each block contains a unique header and is individually identified by its block header hash.

- **PrevHash:** An encrypted hash based on the data contained within the block (including the previous block's hash).
- **BlockHash:** The primary identifier of a block is its cryptographic hash, a digital fingerprint made by hashing the block header twice through the SHA256 algorithm. The resulting 32-byte hash is called the block hash.
- **Timestamp:** A string of characters that uniquely identifies the block and indicates when it was added to the chain is called a timestamp.
- **Merkle Tree:** A Merkle Tree stores all the transactions in a block by producing a digital hash of the entire set of transactions. It enables the user to verify whether or not a particular transaction can be added to a block.
- **BlockBody:** This part contains all the transactions in the block hashed into a Merkle tree. The Merkle root, which is stored in the block body, is the root hash of a Merkle tree. BlockBody stores all transactions and their hashes.

---

**Algorithm 1:** ADR newblock event

---

**Input:** valid nodes list, prev block Hash, initial tx
**Output:** block for valid transactions
1:  **procedure** newblock(valid_node, new_event1)
2:      blocks1.append(new_event1.payload1)
3:      tip = dag_head[block_length]. append(hash(new_event1.payload1))
4:      hash = joining(hash(new_event1. payload1))
5:      **nodeList** ←ADR consensus algorithm
6:      Broadcast(new_event1.payload1)
7:      ts = time.now()
8:      new ← new + random.exp ()
9:      Generate next block event at new_event1
        generateNewBlock (tm, hash, tip)
10:     **return** block_length
11: **end procedure**

---

### 3.4 Block Mining and Ordering

In traditional Proof of Work blockchains, some miners use a lot of processing power to secure the ledger. After solving the computational puzzles, miners are rewarded with coins. Nodes are also present in our ADR protocol. The data format of the transaction record is what this research fundamentally changes. The main difference in the reward structure of ADR compared to a traditional blockchain is that miners that confirm blocks get the reward in the form of an increased ranking score, which makes them a stable and reliable node in the system. It incentivizes the nodes to reference all unreferenced blocks. The leaves of the ADR are newly formed blocks that any other blocks have not yet referenced. In ADR protocol, fewer nodes need to join mining pools, increasing mining decentralization, while more blocks must be mined overall with a lower difficulty. The ranking system is also adjusted to reflect the change in the data structure from a linear blockchain to a ranked directed acyclic graph. It results in nodes getting a ranking not only for the miner who solves the block but for several other nodes who confirm it. The more confirmations a block in the consensus has, the higher the probability that it is a valid block out of two conflicting ones.

---

**Algorithm 2**  Block ordering

---

**Input:** S-Layer list, new_block1
**Output:** An ordered list of secured layer confirmed and unconfirmed blocks
1. **procedure** BlockOrdering
2.          Initialize the block list
3.          todo_p.push(genesisBlock)
4.          **while**_block_p == 0 do
5.                     B ← todo_p.pop()
                       **If**(orderlist(n) == u){
6.                     orderList_conf.add(B)
7.                     child(B) ← {j:(j,B) ∈ E)
8.                     Sort(child(B)) based on their hash values
                             }
9.                     **else**
10.                              todo_p.push(C)
                                 ordListnon_conf.add(B)
11.                     **end if**
12.              **end while**
13.              **return ordList**
14. **end procedure**

---

### 3.5  Security Layer

Layers in the DAG directly affect the reputation of the nodes while ranking the growth evaluation. There are two types of blocks in the system, called confirmed and unconfirmed blocks, as shown in Fig. 2. A graph is divided into security layers named Secured Layer (S-Layer) and Unsecured Layer (U-Layer). Every transaction that's sent flows into the memory pool (mempool) before nodes can confirm it. There may be a spike in transaction activity, which may lead to the congestion of the mempool due to the high number of transactions waiting to be allocated in the next block.

### 3.6  Node Ranking Protocol

The ADR consensus protocol based on the ranking value of nodes is composed of ranking value evaluation, ranking value growth rate evaluation, and the Session Control Protocol (SCP). In the ADR node-set composed of all nodes, firstly, the ranking value evaluation algorithm is used to evaluate the ranking value and growth rate based on several node indicators to ensure the high-quality level of the consensus node-set. Secondly, selecting nodes randomly among nodes with the voting authority ensures the consensus node's randomness and fairness to produce blocks and verify transactions. Thirdly, start the consensus process according to the optimized ADR protocol to ensure the validity and scalability of the consensus. Finally, for the joining or exiting node's request in the system, the joining and exiting of the node are completed according to the Session Control Protocol (SCP). It enhances the dynamics of the blockchain system. The nodeRank mechanism is given in Fig. 3, which describes the steps for the ranking mechanism in nodeRank. In general, the ranking-based consensus efficiently improves the nodes' reliability and system security and surely enhances the scalability of a blockchain network. Our concept of ranking value comprises four elements: ranking points, resources occupied, the ratio of faulty nodes, and the acyclic graph layer level. A node or wallet may govern addresses. Nodes in the ADR protocol must be registered with official addresses (accounts) for ranking, rating, and voting rounds. Instead of pure stake voting, a ranking score governs the nodes' participation according to the rank of an account. Let's describe our model as follows:

Node weight = Rank score = $R_k$ = f (P, U, R)

$$Rk = C_1 P + C_2 U + C_3 R \tag{2}$$

A nodeRank mechanism is used for the implementation of the ADR ranking module. Overall node rank can be calculated by Eq. (2), in which $R_k$ stands for node rank, which is comprised of the sum of P for private key authentication, U for resources occupied, and R for the current rank, where $c_1$, $c_2$, and $c_3$ are the constants. The rank of each node address is determined recursively by the ranks of other addresses linked in the initial nodeRank mechanism. Initially, all addresses receive first rank scores with an equal probability of N (where N is defined as the number of nodes); after that, rank scores are updated using the mathematical Eq. (3):

$$N_R = \frac{i\left(\dfrac{FR_1}{OL_1} + \cdots + \dfrac{FR_n}{OL_n}\right)}{(1-i)} \tag{3}$$

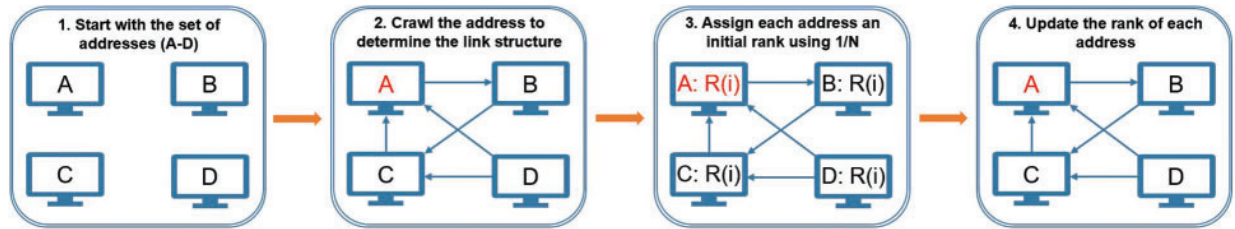

**Figure 3:** Overall node rank mechanism

where $N_R$ denotes the rank of node A, $FR_n$ indicates the rank of the subsequent node linked to A, $OL_n$ represents the number of outbound links on address $FR_n$, and i denotes the initial rank factor such that $0 \le i \le 1$. It should be noted that the $FR_n$ associated with addressing A does not evenly affect the node rank of A. With the Noderank algorithm, the node rank of an address is impacted by the number of outbound links $OL_n$ on an address B, which means that the lower A's node ranking, the less O(Ln) A has to B. The weighted node rank of address B is then added together. The resultant effect of an additional inbound link for address A always increases address A's node ranking.

Overall, an initial rank factor is multiplied by the total weighted ranks of all addresses. The proposed node rank mechanism is shown in Fig. 3, where the ADR system checks for valid node keys and computes the rank of the new node. Here we pass the node to the newRank function, which ultimately finds a random rank for a new node, which is given as:

I = rand(n) where rand array [] is >0.1 and <0.5.

In this case, the total time is considered in seconds with a 440-second target spacing. As a result, we can verify and validate the transaction in this time frame. The difficulty level is calculated by using the hashes of the ADR's block. A malicious actor could use the proposed protocol's flaws to their advantage. Any system that can withstand arbitrary errors can also withstand arbitrary malicious actors. Malicious actors are expected to collide with other nodes and create bogus link structures.

However, the node rank mechanism may produce node ranks with the same value for two nodes. To avoid similarities between two nodes, the $F_R$ in Eq. (4) is rewritten as:

$$FR \sim = \sum_{z=1}^{N} (N_R) \tag{4}$$

After getting these parameters, the rank is updated for the nodes actively participating in the block production and transaction verification consensus. This rank update is a reward for the nodes being true in the system and actively verifying the system mining process. All the notations are mentioned in Table 1.

**Table 1:** Notation table

| Attributes | Mark | Attributes | Mark |
|---|---|---|---|
| Node's ID | *** | Node specific rank | $N_R$ |
| Node's public key | PK | Total number of consensus participation | M |
| Ranking list | $R_k = [r_1, r_2, \ldots, r_j]$ | rand(n) random number | I |
| Number of outbound links on the address | $OL_n$ | Malicious nodes in consensus | $N_m$ |
| The rank of the next node that is linked to A | $FR_n$ | True nodes in consensus | $N_{tr}$ |
| Time level of joining the network | Ti | List of processed transaction sizes | $T_{Xn} = [T_{X1}, T_{X2}, \ldots, T_{Xn}]$ |

---

**Algorithm 3:** Node ranking and updating

---

**Input:** New node, node array
**output:** Ranked node-set, updated node-set
1. **procedure** RankNodes(valid node, node key)
2.        if key == true then
3.               if nodes == "new" then
4.                      RankNewNode(node);
5.               else if(verifyRankNode(node)!<0);
6.                      FR = getForward(node);
7.                      OL = getoutBack(node);
8.                      calculates the new rank
9.               else
10.                     sessionProtocol(node);
11.        else
12.               sessionProtocol(node);

---

### 3.7 Liveness and Fault Tolerance of the System

Liveness for distributed systems is defined as a protocol's ability to communicate among nodes and have the nodes successfully reach a consensus. It indicates that a system's liveness is assured. Instead of focusing on the messages themselves, blockchains frequently attempt to establish which history of transactions is accurate. The termination of a distributed computation is one of many instances of liveness. Liveness is the guarantee that all validators come to some agreement on a value. Our investigation shows that the designated consensus algorithm is live and can defend against malicious miner DoS attacks. Digital signatures demonstrate the authenticity and non-repudiation of communication messages passing across the system. Failure happens when the system cannot provide the required service due to faults; any wrong step in a process is known as a fault, and an error is a tangible measure of the variation from expected behavior that fails. As we have given in Eq. (5), $N_m$ indicates the set of malicious nodes, while $N_T = N_{tr} \cup N_m$. According to our consensus protocol, $\theta$ is defined as a threshold. In the ADR model, $\theta \leq \dfrac{1}{3}$ is required to ensure the liveness of the consensus network.

$$\frac{N_m + N_{tr}}{N_T} \tag{5}$$

ADR protocol provides greater consistency by identifying inconsistencies with a probability of over 99% throughout every procedure. It indicates that there is only a less than 1% chance of neglecting inconsistencies. ADR involves validating the hash values of the initial world states from ($\lfloor (1 - 0)|NT| \rfloor$ + 1) nodes throughout the initialization process.

Based on the honesty assumption, a node only needs to be initialized with consistent information. Inconsistencies can be identified by constant checking, which can be performed for data writing and data reading procedures. In order to complete the consistency verification process, all nodes must gather hash values of the current ledger state that have persisted from at least 5 nodes.

### 3.8 Session Control Protocol (SCP)

The SCP ensures that malicious nodes do not significantly affect the system. It detects malicious nodes and assigns exit tickets to those nodes to leave the consensus. Moreover, the system becomes operational as it is, without any crashes. Even after the node leaves or joins the system, condition (n-1)/3 can still be met if the system remains unaffected.

#### 3.8.1 Entry Ticket Request

In order to join the consensus, new consensus nodes send their entry requests to the SCP in the form of Eq. (6).

$$R_{en} = (K_{Pb}, K_{Pr}, N_R) \tag{6}$$

$R_{en}$, the entry request from the node, is first verified by its public and private keys, and the default ranking $N_R$ is to be added. The primary node verifies and then broadcasts the "node entry confirmation message" to the network. The confirmation message is defined in the form of Eq. (7):

$$\text{PrepareRequest} = ((R_{en}), \text{sign}) \tag{7}$$

#### 3.8.2 Leave Ticket Request

The leave ticket for nodes is issued by the SCP when the reputation of some nodes becomes very low, which can lead to malicious activity or affect the liveness of the network. The other case for the left ticket is when the node itself wants to exit the consensus and no longer wants to be involved in the consensus process. The request can be given as follows using Eq. (8):

$$R_L = (K_{Pb}, K_{Pr}, I_R) \tag{8}$$

After nodes exit, if the total number of consensus nodes reduces to 3f + 1, we start the node dynamic joining phase. Otherwise, secondary nodes receive exit confirmation messages signed by both the primary node and the node to be exited.

$$\text{PrepairJoinRequest} = ((\text{RL}), \text{sign}) \tag{9}$$

Based on the explanations in the above sections, the overall process of the ADR protocol is shown in Figure 4.

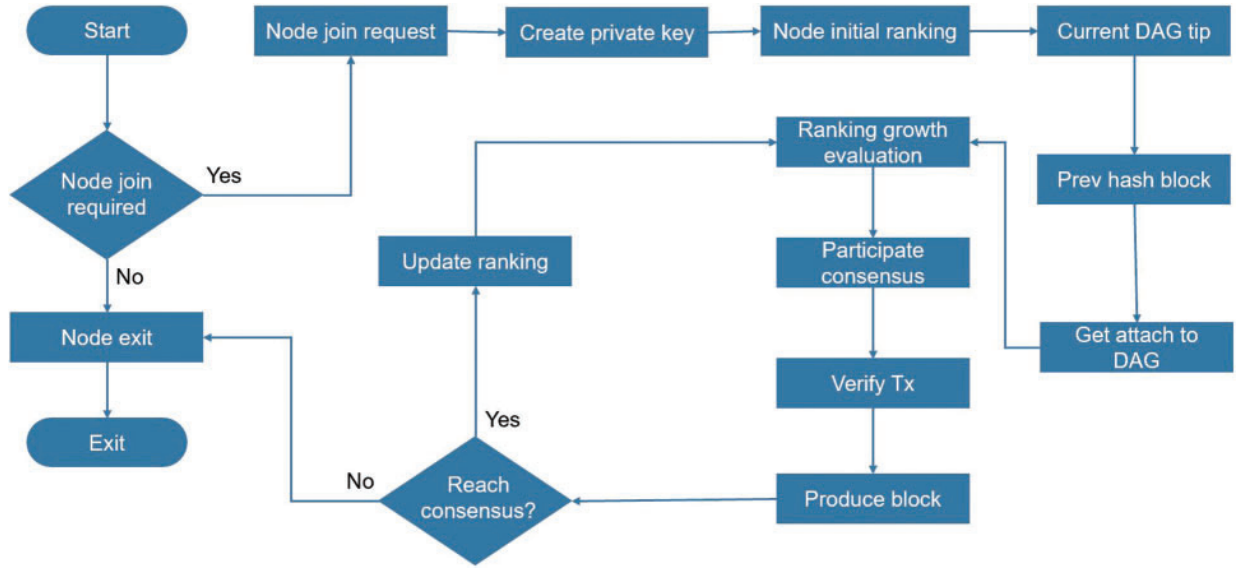

**Figure 4:** Overall flow chart of ADR protocol

## 4 Simulation Results

In order to assess our design, we used a blockchain framework with Golang to create a working prototype of an ADR blockchain system that can serve as a consensus protocol for IoT networks. Tests were run on the original IOTA, ByteBall, and ADR to compare them to IOTA and ByteBall. Performance metrics and system network overhead experimental results are shown in this section. Various experimental schemes were created and tested on our cluster server machines, each of which has an Ubuntu 16.04 operating system and a two-core Intel Core i5 2.4 GHz CPU with 8 GB of RAM and a 256 GB SSD. It allowed us to confirm the applicability of the ADR consensus protocol. A framework was set up to build simulation networks with the primary nodes in their default positions. The maximum number is 40, and the initial consensus committee comprises 10 simulated machines. In the performance analysis, the primary indicators examined were transaction latency, throughput, communication times, election fairness, epochs while the system was running, and how malicious nodes affected the system's scalability.

### 4.1 Throughput

Transaction throughput (TPS) is a crucial metric to measure the concurrency and scalability of the blockchain system. The efficiency of the consensus mechanism and the capacity to handle transactions increase with throughput scalability. The system's throughput increased as more nodes joined the consensus, and the higher throughput helped us achieve better results than other blockchains. Eq. (10) allowed us to determine the throughput scalability of the system.

$$T = T_b/N_t \tag{10}$$

where $T$ represents throughput, $T_b$ is block time, and $N_t$ is the total number of transactions during that time. This experiment evaluates the throughput value by setting the average transactional values for each block. The client continues to send transaction data to the system. About 20 blocks were used to test the throughput with varying numbers of transactions. Findings revealed that when more nodes were added, the consensus capacity to manage an increasing volume of transactions increased. The comparative results in Fig. 5, indicate significant outcomes compared to ByteBall and IOTA. The presumption of consistency mainly influenced consensus. Since the complexity of reaching consensus increased with the committee's size, tight consistency significantly slowed down performance. As a result of its fork tolerance, poor consistency, on the other hand, can substantially scale the blockchain. Consensus protocol architecture, the percentage of malicious nodes, and other aspects are also important. A system's performance can be more precisely assessed by counting its total TPS, multiplying each chain's throughput, and counting participating nodes. It was noted that current DAG blockchain technologies could not accomplish both simultaneously. Assume that these two variables fall on either side of the spectrum. When weighing scalability versus throughput, it is unclear what the best option is for the sweet spot. Physical boundaries or security concerns could place restrictions on the bottleneck.

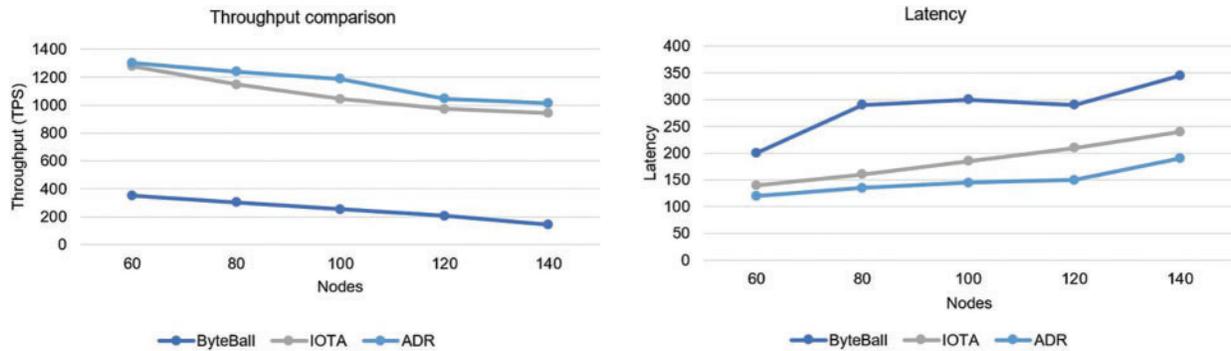

**Figure 5:** Throughput and latency comparison

### 4.2 Latency

Transaction latency, defined as the period between the submission of a transaction and its confirmation, serves as the standard for assessing the communication efficiency and operational duration of the blockchain network's consensus mechanism. However, there would be less latency, which would speed up transaction confirmation and lessen the chance of the blockchain fragmentation phenomenon, with the evaluation of the ADR protocol, ByteBall, and IOTA algorithms' transaction latency times at the following nodes: 50, 70, 90, 110, 130, and 150. As shown in Fig. 5, it can be seen that the blocks were chosen randomly and with different numbers of transactions. Each block underwent 15 tests. Each block's average latency was then compared to various algorithms. As can be seen, the ADR protocol's latency was substantially lower than IOTA's ByteBall algorithm, demonstrating that the node's consensus speed was faster than other nodes. Additionally, latency varies depending on the number of transactions in a block. Network propagation and node-to-node verification take longer when more transactions are in larger blocks. Given the constrained network bandwidth, the broadcast confirmation delay increased with increasing latency.

### 4.3 Fault Tolerance Examination

The experimental investigation of the consensus performance is done because the ADR consensus protocol includes a reputation-based mechanism. There are two sections in this experiment:

### 4.3.1 Fair Node Election

First, 20 nodes are created with the ADR protocol, and their reputations are set to 0.5 by default. The odd number of nodes labeled as "faulty nodes" and the even number labeled as "true nodes" are shown in Fig. 6. We ran 50 experiments to give each node a fair chance to be evaluated. To determine the impact on the consensus protocol for IOTA, Phantom, and ADR, we measured the throughput by adding faulty nodes to the system by the n-1/3 factor. The experiment shows better results and higher liveness of the network for the ADR consensus protocol, which is a significant signal of improvement in the consensus protocol testing.

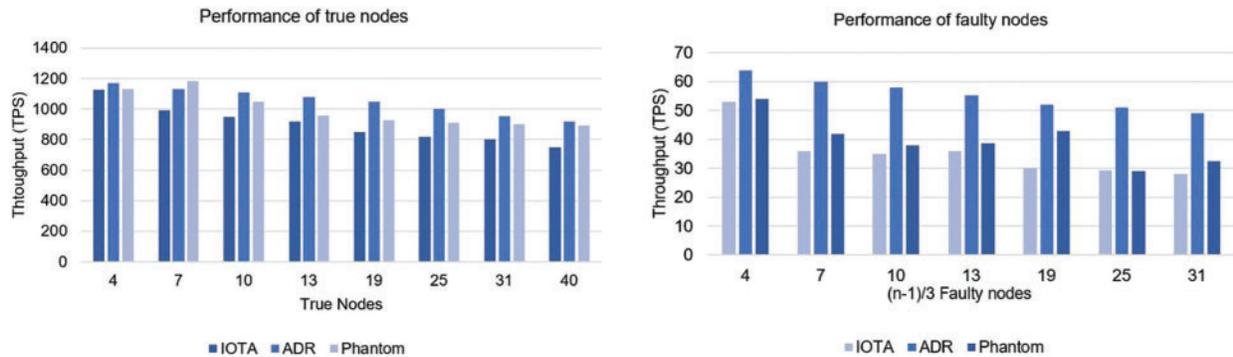

**Figure 6:** Performance of true and faulty nodes

### 4.3.2 Activity of Malicious Nodes

In this section, we evaluated the throughput by adding faulty nodes to the system by the n-1/3 factor to see the effect on the consensus mechanisms for IOTA, Phantom, and ADR. Fig. 6 shows the decline in throughput with the faulty nodes' addition, which affects the system's scalability. In contrast, ADR still shows good results, having higher throughput than other protocols.

When the true nodes were run on the system, no faulty nodes were added, the system's scalability and throughput increased, and ADR outperformed IOTA and Phantom.

### 4.4 Epoch Transition and Block Time

This section demonstrates how the ADR can benefit from throughput improvements. It was tested whether the reputation ratings were relevant to their capabilities by deploying various validators with varying capabilities. The network can be scaled thanks to the significant improvement in epoch throughput, as seen in Fig. 7. To be more precise, the node's capability was set to 100% and restricted the speed at which the validators may process transactions. As a result, a validator with a low capability provides more unreliable results than one with a high capability. In this scenario, 100 nodes were used, and 10 blocks with various transactions were made to compare with IOTA and Phantom.

Furthermore, each validator controls the variables honestly, and the range of their skills (from 5% to 100%) is uniform. The benefits of the ADR protocol for throughput enhancement can be seen in Fig. 7. Due to the poor capability leader at first, the throughput was less than 1300 TPS. However, due to the choice of a more capable leader based on its cumulative ranking score after roughly 8 epochs, the throughput has stabilized at an average of 1600 TPS. 20 iterations of the system's average throughput were also performed, resulting in 900 TPS for IOTA, ByteBall, and Phantom. An epoch test with 200 shards and 100 nodes was conducted. Here, the block production time of the ADR consensus was also added. Block time is a unit of measurement for how long it takes a network's miners or validators to confirm a block's transactions and create a new block in that blockchain. Fig. 7 gives the ADR production time, which was relatively better and working well as the miners were true and no faulty nodes were added. The simple attack performed poorly for the first two epochs. Still, it took the malicious node a significant amount of time (about 10 epochs) to accumulate enough rank points to be chosen as the leader again. It indicated that the system could continue to function effectively. It can be inferred that the ADR can improve functionality and security when dealing with hostile node attacks.

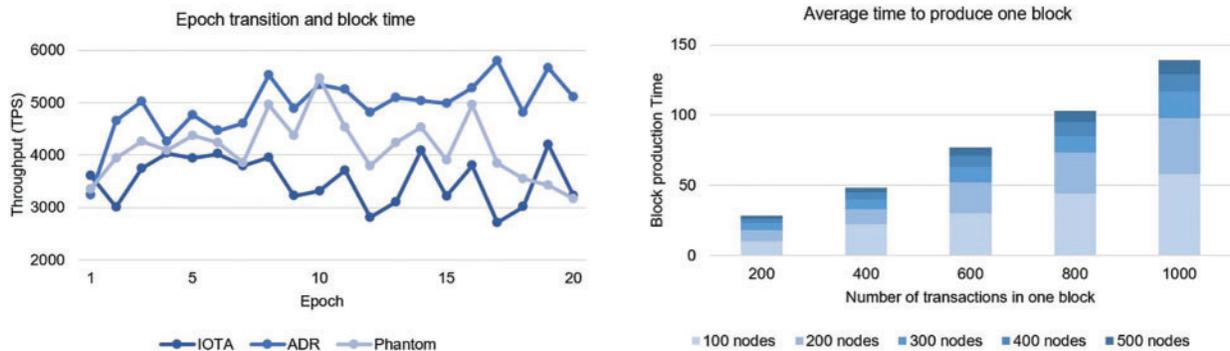

**Figure 7:** Epoch testing and ADR production time

### 4.5 Implications

According to the findings of this research, the ADR protocol performed better than IOTA and ByteBall in terms of throughput and latency for consensus methods. Furthermore, data analysis demonstrated that the consensus technique would, to some extent, increase delays and decrease throughput, resulting in a performance bottleneck. This finding has been verified on both IOTA and ByteBall. The consensus algorithm of the blockchain essentially struck a balance between fault tolerance and security. Additionally, the examination of this paper demonstrated that under various transaction quantities, outcomes were improved in terms of latency and throughput. IOTA and ByteBall's latency drastically rose as the number of transactions rose, while ADR's throughput underwent considerable variations. The executive and data layer operations were set to be as minimal as possible in the experimental circumstances. Thus, the consensus mechanism mainly determined how well the blockchain performed overall. In other words, the performance of these blockchain platforms under various transaction volumes reflected the performance of the consensus algorithms. The selection of consensus algorithms or other private blockchain platforms impacted the throughput and latency of blockchains throughout the development process. The number of transactions was also quite significant when choosing a blockchain platform and somewhat influenced the performance measures. When choosing a blockchain platform, the number of transactions was also a significant factor and somewhat influenced the performance measures.

### 4.6 Limitations

The performance of the consensus algorithms in private blockchains is the main topic of this research paper. The throughput scalability, network delay, and liveness of the consensus were somewhat coupled because of the design of the blockchain itself, making it impossible for them to be autonomous. The primary goal of this work was to develop a consensus mechanism that maximizes throughput and scalability while minimizing latency and delay in block formation and verification, followed by a performance analysis of the blockchain. The decoupling between throughput and scalability may perform better with the blockchain's ongoing development. The experiment in this paper assessed the performance of IOTA and ByteBall's consensus algorithms in scenarios involving multiple servers. Numerous distributed consensus systems running on blockchain data structures are currently being suggested. The experiment in this paper assessed the performance of general blockchain frameworks like Bitcoin, Ethereum, and Hyperledger Fabric and DAG-related consensus mechanisms such as IOTA, ByteBall, and Phantom consensus algorithms in scenarios involving multiple servers. The study's findings suggest that more research is needed to determine when the number of nodes varies and how the number of transactions affects the performance of the consensus algorithms used by blockchain platforms. The comparison of ADR with other blockchain protocols is shown in Table 2.

**Table 2:** Comparison of ADR with other blockchain protocols

| Serial number | Features | Bitcoin | Ethereum | Hyperledger fabric | IOTA | ByteBall | Spectre | Phantom | ADR |
|---|---|---|---|---|---|---|---|---|---|
| 1. | Fully developed | ✓ | ✓ | ✓ | Development mode | Development mode | ✓ | ✓ | Development mode |
| 2. | Blockchain type | Public | Public, private, hybrid | Private | Public | Public | Public | Public | Private |
| 3. | Trustless | ✓ | ✓ | Trusted validator nodes | ✓ | Rely on honest witnesses | Honest participants | Honest nodes | Trusted validator nodes |
| 4. | Multiple applications | Financial only | ✓ | ✓ | Financial only | Financial only | ✓ | ✓ | Private blockchain |
| 5. | Consensus algorithm | PoW | PoW, PoS | PBFT | Tip selection algorithm | Tip selection algorithm | Block voting algorithm | Greedy selection algorithm | ADR |
| 6. | Fee less | × | × | Optional | ✓ | ✓ | × | × | ✓ |
| 7. | Tx integrity and authentication | ✓ | ✓ | ✓ | ✓ | ✓ | ✓ | ✓ | ✓ |
| 8. | Data confidentiality | × | × | ✓ | × | ✓ | ✓ | × | ✓ |
| 9. | ID management | × | × | ✓ | × | ✓ | × | × | ✓ |
| 10. | Key management | × | × | (through CA) | × | × | × | × | ✓ |
| 11. | User authentication | Digital signature | Digital signature | Based on the enrolment certificate | Digital signature | Attestation | Pairwise voting procedure | Pairwise voting procedure | Digital signature |
| 12. | Vulnerability to attack | 51%, linking attacks | 51% | >1/3 faulty nodes | 34% attack | DOS Attacks | 50%, side channel attacks | >50% | Up to 9% |
| 13. | Tx throughput | 7 TPS | 8–9 TPS | It depends on the network size | 850 TPS | 15 TPS | NA | NA | 1250 TPS |
| 14. | Latency | 10 mins | 15–20 secs | Less than Bitcoin and Ethereum | 200 sec | 30 sec | 1–10 sec | Fast than Spectre | 50 sec |
| 15. | Scalability | × | × | × | ✓ | ✓ | ✓ | ✓ | ✓ |

## 5 Conclusion

Private blockchain systems' performance and scalability have become increasingly critical as blockchain business applications have gained popularity. This research demonstrated that the performance of the ADR consensus protocol is highly scalable, fast, and secure. The proposed ADR protocol used ranking-based node election and a dynamic Session Control Protocol (SCP) among nodes to achieve significant consensus performance, low network latency, and excellent scalability. High-ranked nodes were randomly selected to join the SCP consensus safety algorithm. The simulation results demonstrated that the ADR protocol outperformed the ByteBall and IOTA algorithms in average throughput and scalability tests while considering security measures. Finally, experiments were conducted on EC2 clusters to show a promising solution for the performance bottleneck of traditional consensus mechanisms and a realistic solution for integrating blockchain technologies into a private business application domain. Hence, this study is essential for future research and algorithm optimization of the consensus mechanism.

**Funding Statement:** The researchers did not receive a specific grant from public, commercial, or not-for-profit funding agencies.

**Conflicts of Interest:** The authors declare they have no conflicts of interest to report regarding the present study.

## References

[1]  Y. Tian, C. Li, X. Chen and J. Li, "An efficient anti-quantum lattice-based blind signature for blockchain-enabled systems," *Information Sciences*, vol. 546, no. 2, pp. 253–264, 2021.

[2]  C. Li, M. Dong, J. Li, G. Xu, X. Chen *et al.,* "Healthchain: Secure EMRs management and trading in distributed healthcare service system," *IEEE Internet of Things Journal*, vol. 8, no. 9, pp. 7192–7202, 2021.

[3]  M. Dong, C. Li, J. Li, G. Xu, X. Chen *et al.,* "Efficient medical big data management with keyword-searchable encryption in healthchain," *IEEE Systems Journal*, vol. 16, no. 4, pp. 5521–5532, 2022.

[4]  S. Zhang and J. H. Lee, "Analysis of the main consensus protocols of blockchain," *ICT Express*, vol. 6, no. 2, pp. 93–97, 2020.

[5]  H. Kalodner, M. Carlsten, P. Ellenbogen, J. Bonneau and A. Narayanan, "An empirical study of Namecoin and lessons for decentralized namespace design," in *14th Annual WEIS*. Delft, The Netherlands, 1–21, 2015.

[6]  M. Carlsten, H. Kalodner, A. Narayanan and S. M. Weinberg, "On the instability of Bitcoin without the block reward," in *Proc. ACM Conf. Computer Communication and Security*, Vienna, Austria, pp. 154–167, 2016.

[7]  S. Secinaro, D. Calandra and P. Biancone, "Blockchain, trust, and trust accounting: Can blockchain technology substitute trust created by intermediaries in trust accounting? A theoretical examination," *International Journal of Management Practice*, vol. 14, no. 2, pp. 129–145, 2021.

[8]  J. Yadav and R. Shevkar, "Performance-based analysis of blockchain scalability metric," *Tehnički glasnik*, vol. 15, no. 1, pp. 133–142, 2021.

[9]  M. Castro and B. Liskov, "Practical Byzantine fault tolerance and proactive recovery," *ACM Transactions on Computer Systems (TOCS)*, vol. 20, no. 4, pp. 398–461, 2002.

[10]  F. R. Yu, J. Liu, Y. He, P. Si and Y. Zhang, "Virtualization for distributed ledger technology (vDLT)," *IEEE Access*, vol. 6, pp. 25019–25028, 2018.

[11]  S. Popov, O. Saa and P. Finardi, "Equilibria in the tangle," *Computers & Industrial Engineering*, vol. 136, no. 2, pp. 160–172, 2019.

[12]  K. Gai, Z. Hu, L. Zhu, R. Wang and Z. Zhang, "Blockchain meets DAG: A BlockDAG consensus mechanism," in *Proc. ICA3PP*, New York, USA, pp. 110–125, 2020.

[13] Y. Sompolinsky, S. Wyborski and A. Zohar, "PHANTOM GHOSTDAG: A scalable generalization of Nakamoto consensus," in *Proc. of AFT '21: 3rd ACM Conf. on Advances in Financial Technologies*, Arlington, VA, USA, pp. 57–70, 2021.

[14] C. Yu, W. Yang, F. Xie and J. He, "Technology and security analysis of cryptocurrency based on blockchain," *Complexity*, vol. 2022, no. 2, pp. 1–15, 2022.

[15] F. Gai, B. Wang, W. Deng and W. Peng, *Proof of Reputation: A Reputation-based Consensus Protocol For Peer-to-Peer Network*. Gold Coast, QLD, Australia: Springer, 2018. [Online]. Available at: https://link. springer.com/chapter/10.1007/978-3-319-91458-9_41.

[16] D. Shrier, W. Wu and A. Pentland, "Blockchain & infrastructure (identity, data security)," *Massachusetts Institute of Technology-Connection Science*, vol. 1, no. 3, pp. 1–19, 2016.

[17] Q. Zhou, H. Huang, Z. Zheng and J. Bian, "Solutions to scalability of blockchain: A survey," *IEEE Access*, vol. 8, pp. 16440–16455, 2020.

[18] K. Croman, C. Decker, I. Eyal, A. E. Gencer, A. Juels *et al.,* "On scaling decentralized blockchains," *in proc. Int. Conf. on Financial Cryptography and Data Security*, Christ Church, Barbados, pp. 106–125, 2016.

[19] M. Kuzlu, M. Pipattanasomporn, L. Gurses and S. Rahman, "Performance analysis of a hyperledger fabric blockchain framework: Throughput, latency and scalability," in *Proc. 2nd IEEE Int. Conf. on Blockchain*, Atlanta, GA, USA, pp. 536–540, 2019.

[20] M. A. Javarone and C. S. Wright, "From Bitcoin to Bitcoin cash: A network analysis," in *Proc. 1st Workshop on Cryptocurrencies and Blockchains for Distributed Systems*, Munich, Germany, pp. 77–81, 2018.

[21] M. A. Imtiaz, D. Starobinski, A. Trachtenberg and N. Younis, "Churn in the bitcoin network: Characterization and impact," in *Proc. IEEE Int. Conf. Blockchain Cryptocurrency*, Seoul, South Korea, pp. 431–439, 2019.

[22] L. Luu, V. Narayanan, C. Zheng, K. Baweja, S. Gilbert *et al.,* "A secure sharding protocol for open blockchains," in *Proc. ACM SIGSAC Conf. on Computer and Communications Security*, Vienna, Austria, pp. 17–30, 2016.

[23] E. Kokoris-Kogias, P. Jovanovic, L. Gasser, N. Gailly, E. Syta *et al.,* "OmniLedger: A secure, scale-out, decentralized ledger via sharding," in *Proc. IEEE Symp. on Security and Privacy (SP)*, San Francisco, USA, pp. 583–598, 2018.

[24] M. Zamani, M. Movahedi and M. Raykova, "RapidChain: Scaling blockchain via full sharding," in *Proc. ACM Conf. on Computer and Communications Security*, Arlington, VA, USA, pp. 931–948, 2018.

[25] J. Wang and H. Wang, "Monoxide: Scale out blockchain with asynchronous consensus zones," in *Proc. 16th USENIX Symp. on Networked Systems Design and Implementation*, Boston, MA, USA, pp. 95–112, 2019.

[26] I. Eyal, A. E. Gencer, E. G. Sirer and R. Van Renesse, "Bitcoin-NG: A scalable blockchain protocol," in *Proc. 13th USENIX Symp. on Networked Systems Design and Implementation*, Santa Clara, CA, USA, pp. 45–59, 2016.

[27] C. T. Nguyen, D. T. Hoang, D. N. Nguyen, D. Niyato and H. T. Nguyen *et al.,* "Proof-of-stake consensus mechanisms for future blockchain networks: Fundamentals, applications and opportunities," *IEEE Access*, vol. 7, pp. 85727–85745, 2019.

[28] X. Yuan, F. Luo, M. Z. Haider, Z. Chen and Y. Li, "Efficient byzantine consensus mechanism based on reputation in IoT blockchain," *Wireless Communication and Mobile Computing*, vol. 2021, no. 6, pp. 1–17, 2021.

[29] J. H. Lin, K. Primicerio, T. Squartini, C. Decker and C. J. Tessone, "Lightning network: A second path towards centralisation of the Bitcoin economy," *New Journal of Physics*, vol. 22, no. 8, pp. 083022, 2020.

[30] Y. Buchnik and R. Friedman, "FireLedger: A high throughput blockchain consensus protocol," *VLDB Endowment*, vol. 13, no. 9, pp. 1525–1539, 2020.

[31] C. S. A. Worley, "Blockchain Tradeoffs and Challenges for Current and Emerging Applications: Generalization, Fragmentation, Sidechains, and Scalability," in *Proc. IEEE, 2018 Int. conf. on Internet of Things (iThings) and IEEE Green Computing and Communications (GreenCom) and IEEE Cyber, Physical and*

      *Social Computing (CPSCom) and IEEE Smart Data (SmartData)*, Halifax, NS, Canada, pp. 1582–1587, 2018.

[32] B. Cao, Z. Zhang, D. Feng, S. Zhang, L. Zhang *et al.,* "Performance analysis and comparison of PoW, PoS and DAG based blockchains," *Digital Communication and Networks*, vol. 6, no. 4, pp. 480–485, 2020.

[33] Y. Lewenberg, Y. Sompolinsky and A. Zohar, *Inclusive Block Chain Protocols*. San Juan, Puerto Rico: Springer, 2015. [Online]. Available at: https://link.springer.com/chapter/10.1007/978-3-662-47854-7_33.

[34] C. Fan, S. Ghaemi, H. Khazaei, Y. Chen and P. Musilek, "Performance analysis of the IOTA DAG-based distributed ledger," *ACM Transactions on Modeling and Performance Evaluation of Computing Systems*, vol. 6, no. 3, pp. 1–20, 2021.

[35] A. Churyumov, *Byteball: A Decentralized System for Storage and Transfer of Value*. New York, NY, USA: Byteball, 2016. [Online]. Available at: https://obyte.org/Byteball.pdf.

[36] C. Li, P. Li, D. Zhou, Z. Yang, M. Wu *et al.,* "A decentralized blockchain with high throughput and fast confirmation," in *in proc. USENIX Annual Technical Conf.*, Berkeley, CA, USA, pp. 515–528, 2020.